\documentclass{statsoc}

\usepackage[a4paper]{geometry}
\usepackage{graphicx}
\usepackage{amsmath,amssymb}
\usepackage[utf8]{inputenc}
\usepackage{natbib}

\title[Ensemble Prediction for CD]{Ensemble Prediction of Time to Event Outcomes with Competing Risks: A Case Study of Surgical Complications in Crohn's Disease}
\author{Michael C Sachs}
\address{Department of Medicine, Karolinska Institutet
Stockholm,
Sweden.}
\email{michael.sachs@ki.se}
\author{Andrea Discacciati}
\address{Institute of Environmental Medicine, Karolinska Institutet,
Stockholm, Sweden.}
\author{Åsa H Everhov}
\address{Department of Medicine, Karolinska Institutet,
Stockholm, Sweden.}
\author{Ola Olén}
\address{Department of Medicine, Karolinska Institutet,
Stockholm, Sweden.}
\author[Michael C Sachs {\it et al.}]{Erin E Gabriel}
\address{Department of Medical Epidemiology and Biostatistics, Karolinska Institutet,
Stockholm, Sweden.}

\begin{document}
  
\begin{abstract} We develop a novel algorithm to predict the occurrence of major abdominal surgery within 5 years following Crohn's disease diagnosis using a panel of 29 baseline covariates from the Swedish population registers. We model pseudo-observations based on the Aalen-Johansen estimator of the cause-specific cumulative incidence with an ensemble of modern machine learning approaches. Pseudo-observation pre-processing easily extends all existing or new machine learning procedures to right-censored event history data. We propose pseudo-observation based estimators for the area under the time varying ROC curve, for optimizing the ensemble, and the predictiveness curve, for evaluating and summarizing predictive performance. 
\end{abstract}

\section{Introduction}
\label{sec:intro}

\subsection{Motivating study and statistical approaches}
Crohn's disease (CD) is a chronic debilitating condition characterized by periods of inflammatory activity in the bowel that causes symptoms such as abdominal pain, diarrhea, and weight loss. Pharmacologic treatment for CD includes medications such as steroids, immunomodulating drugs, and biological therapy. Despite these available medications, many people with CD are escalated to surgical interventions from small to extensive resections of the bowel or colon \citep{gomollon20163rd}. Previous studies have estimated that up to 50\% of patients with CD undergo surgery within 10 years after diagnosis; however, surgical rates have decreased over time, possibly due to the introduction of modern treatments such as thiopurines and anti-TNF \citep{lakatos2012has,ramadas2010natural}. The aim of this study is to determine whether clinical and demographic characteristics observed at the time of diagnosis can be used to predict the occurrence of major abdominal surgery within 5 years, with the goal of personalized disease management. Although it might be of interest to determine which characteristics are associated with risk, we instead focus on obtaining the most accurate predictions, which, at least in this registry based setting, are useful even without knowing what is driving them.  

One of the statistical challenges is to deal with the fact that patients may die or emigrate during follow-up. These competing risks must be accounted for in development and evaluation of the prediction model. There are several possible approaches available for modeling a censored time to event outcome with competing risks. The time to event can be converted into a binary indicator of having the event within 5 years, and then the remaining observations weighted by the inverse of the probability of not being censored. This can be less efficient as subjects censored prior to the time horizon do not directly contribute to estimation of the model. What's more, the estimation method must be able to incorporate weights. A series of Cox models or direct modeling of the cumulative incidence as a function of covariates is also an option via the Fine-Gray model \citep{fine1999proportional}. 

As it is rarely known which model will give the best predictions, we instead wish to use an ensemble of models and machine learning (ML) procedures for prediction. There are existing methods that adapt ML procedures to censored data. \citet{binder2009boosting} develops a boosting algorithm for the Fine-Gray model under competing risks to select a subset of variables and form a prediction model. \citet{ishwaran2014random} develops a survival random forest that accounts for competing risks. \citet{hothorn2005survival} develop two methods for ensemble learning that deal with right censored data by considering only observed event times and weighting by the probability of not being censored. \citet{scheike2008predicting} describes an alternative approach that involves direct binomial regression of the counting process formulation of the event in the presence of competing risks. \citet{gerds2018} provides a suite of methods for estimating the performance of prediction models in the competing risks setting, using inverse probability of censoring weighted (IPCW) estimates of the Brier score and AUC, however, one must fit the prediction models first. Under many of the existing methods based on binary data, including those in \citet{scheike2008predicting} and \citet{hothorn2005survival}, one needs to model the distribution of time to censoring and then use these as weights in the ML procedure. Although weight estimation may be easy, particularly in our motivating setting, there still needs to be a way to use the IPC weights in the ML procedure, making IPC weighting a useful but procedure specific solution for dealing with censoring. 

Our suggested procedure provides a general method that allows the application of all existing and any newly developed ML algorithm, without extension, to censored data with or without competing risks. We propose to generate pseudo-observations of the cause-specific cumulative incidence of surgery at 5 years, accounting for the competing risks of death and emigration, and use these pseudo-observations as the outcome variable in an ensemble learning approach using standard software for continuous outcomes. \citet{mogensen2013random} takes a similar approach to our proposed method, also using pseudo-observations, but only for the class of tree-based methods, building an ensemble of trees to construct a random forest. One may view our proposed procedure as an extension of concepts in \citet{mogensen2013random}. 

In addition, we demonstrate how pseudo-observations can be used to estimate the time-varying ROC curve \citep{zheng2012evaluating, heagerty2000time} and the time varying predictiveness curve  \citep{pepe2007integrating}. To our knowledge, neither of these results has been demonstrated before. Using these results, we combine a set of ML procedures into an ensemble by optimizing the time-varying area under the ROC curve.  

 The use of pseudo-observations has been described and evaluated fairly extensively for their use in the estimation of association models \citep{andersen2003generalised, andersen2010pseudo}. Relatively little attention has been given towards their use in the development and evaluation of prediction models. In addition to \citet{mogensen2013random}, \citet{nicolaie2013dynamic} develop methods related to the development of dynamic prediction models using pseudo-observations and one paper has suggested evaluating existing predictive models using pseudo-observations to estimate the Brier score \citep{cortese2013comparing}.  

\subsection{Description of the dataset and variables}
The motivating example comes from the Swedish population registers, in which we identified all individuals with a first ever diagnosis of CD in the Inpatient or Outpatient register from January 1, 2003 until December 31, 2014 using the international classification of disease (ICD) code K50 (ICD-10, 1997-2014). We required at least two diagnostic listings with CD as main or contributory diagnosis in inpatient or outpatient care or CD associated abdominal surgery at the same time as the first CD diagnosis. Patients registered with a record of one of the major abdominal surgery events before 1 January 2003 were excluded.

The outcome variable of major abdominal surgery was defined according to a set of procedure codes based on the NOMESCO Classification of Surgical Procedures. We used six categories of procedures based on anatomical localization and expected functional outcome: resection of the small bowel, ileocecal/ileocolic resection, segmental resection of the colon, colectomy, and/or proctectomy. The occurrence of any of those surgeries and possibly in combination with each other defined the major abdominal surgery outcome. Endoscopic, perianal, and upper gastrointestinal procedures were not included. 

Baseline characteristics and demographics are gathered from the patient register, the multigeneration register, and the prescribed drug register which are described in more detail elsewhere \citep{ludvigsson2016registers}. A detailed description of the data sources and methods for defining variables is available in a recent manuscript \citep{olen2017childhood}, and descriptive statistics for our study population are given in the Supplementary Materials.

\subsection{Aims of this paper}
Our aims are four-fold: illustrate our proposed general pseudo-observation based method for ensemble learning optimizing the area under the time varying ROC curve, compare our approach to existing methods, develop prediction algorithms for risk of surgery within 5 years in patients with CD, and evaluate those predictions with the pseudo-observation based predictiveness curve. 

After giving some notation in Section 2.1, we describe pseudo-observations for the cause-specific cumulative incidence in Section 2.2. We then develop pseudo-observation based estimators for the time-varying cause-specific area under the ROC curve and the predictiveness curve in Section 2.3. Then, using these results, we detail our modification of the SuperLearner algorithm for this setting in Section 2.4. All of these methods are evaluated and compared to existing approaches in a simulation study which is described in Section 3. Finally, we analyze and discuss the results from our motivating study in Section 4. 

\section{Methods}

\subsection{Notation}

\begin{itemize}
\item $Y_i$: right censored event time and $\Delta_i \in \{0,1,2, 3\}$: event indicator where 0 indicates censoring, 1 indicates surgery, 2 indicates death, and 3 indicates emigration, for subjects $i = 1, \ldots, n$.
\item $T_i$: is the true event time, $\delta_i \in \{1,2,3\}$: event indicator 1 indicates surgery, 2 indicates death, and 3 indicates emigration, 
\item $X_{ij}$: biomarker measurement/covariate measured at baseline, for $j = 1, \ldots, p$. $\mathbf{X_i}$ is the vector of covariates for subject $i$.
\item $B_{ik} = f_k(\mathbf{X_i})$, where $f_k: \mathbb{R}^p \rightarrow \mathbb{R}$ is a covariate signature indexed by $k$. We denote $\hat{f}_k$ the estimated signature. The index $k$ is used to distinguish between signatures estimated by different algorithms or on different subsets of the data.

We are interested in the cumulative incidence of failure due to surgery at 5 years: 

\[
C_1(t_{\star}) = E\{I(T_i \leq t_{\star}, \delta_i = 1)\} = \int_{0}^{t_{\star}} S(u)\alpha_1(u) \,du,
\]

where $S(u)$ is the overall survival probability, $\alpha_1(u)$ is the cause-specific hazard for failures from abdominal surgery, and $t_{\star} = 5$ years. 

\end{itemize}

\subsection{Pseudo-observations for cumulative incidence}

Following \citet{andersen2010pseudo}, we summarize the survival data as the counting process: 

\[
N_1(t) = \sum_i I(Y_i \leq t, \Delta_i = 1), 
\]
giving the number of observed failures due to surgery on or before time $t$, and $R(t) = \sum_i I(Y_i \geq t)$ gives the number of subjects still at risk just before time $t$. Our estimator of the cumulative incidence function is the Aalen-Johansen estimator \citep{aalen1978empirical}
\[
\hat{C}_1(t) = \int_{0}^{t} \hat{S}(u)\, d\hat{A}_1(u), 
\]
where $\hat{A}_1(u) = \int_{0}^t \, dN_1(u) / R(u)$ is the Nelson-Aalen estimator for the cumulative cause-specific hazard for surgery failures, and $\hat{S}$ is the Kaplan-Meier estimator of the overall survival from any cause. 

Then the $i$th pseudo-observation for cause 1 at time $t$ is 
\[
\hat{C}^i_1(t) = n \hat{C}_1(t) - (n - 1) \hat{C}_1^{-i}(t), 
\]
where $\hat{C}^{-i}_1(t)$ is the cumulative incidence function estimator computed by using the sample excluding the $i$th observation. By construction and the unbiasedness of the Aalen-Johansen estimator, the pseudo-observations are unbiased for the cumulative cause-specific incidence: $E\{\hat{C}^i_1(t)\} = C_1(t).$ Moreover, observe that the survival from any cause can be estimated $E\{1 - \sum_{j=1}^3\hat{C}^i_j(t)\} = P(Y_i > t)$.

\citet{graw2009pseudo} proved that the mean of pseudo-observations conditional on covariates is asymptotically unbiased for the conditional cause-specific cumulative incidence. The validity of this proof and asymptotic properties of pseudo-observations in regression settings was further studied in \citet{overgaard2017asymptotic} and \citet{jacobsen2016note}. Thus we have,

\[
E(\hat{C}^i_1(t_{\star}) | \mathbf{X_i}) = P(T_i \leq t_{\star}, \delta_i = 1 |  \mathbf{X_i}) + o_P(1). 
\]
This property of the pseudo-observations for the cumulative incidence allows us to use them as the outcome in prediction models in which we aim to estimate a mapping $f(\mathbf{X_i})$. Then, as we show in the next section, we can evaluate the mapping $f$ for predicting surgery by estimating the time-varying ROC curve, again using pseudo-observations. 

Conditional asymptotic unbiasedness of the pseudo-observations relies on the assumption that censoring time is independent of event time, event type and all other covariates; this was called `completely independent censoring' in \citet{overgaard2017asymptotic}. In our dataset, due to the quality of the Swedish population register, there is no loss to follow-up that is not due to death or emigration. Thus it is critical to model death and emigration as competing risks rather than simply censoring those subjects because the association between certain covariates and death or emigration may violate this assumption. All other censoring before 5 years is determined by the time of entry into the study (the date of CD diagnosis). For this reason, we use stratified pseudo-observation procedures. This was suggested in \citet{andersen2010pseudo}, as a way to relax the deal with censoring that is dependent on a known categorical covariate. In other studies with loss to follow-up, censoring dependence maybe more complex. Procedures to extend pseudo-observations by modeling the censoring distribution and reweighting have been developed \citep{binder2014pseudo}.

\subsection{Time varying measures of prediction accuracy estimated using pseudo observations}

The time-varying ROC curve \citep{heagerty2000time,saha2010time} for biomarker signature $B_i$ at time $t$ and cutoff $c$ is defined by the time varying, cause-specific true-positive fraction 
\[
TP(t, c) = P(B_i > c | T_i \leq t, \delta_i = 1)
\]
and the false positive fraction that is defined across all event types: 
\[
FP(t, c) = P(B_i > c | T_i > t).
\]
While the $TP$ can be defined for all causes, we omit the cause-type subscript for the $TP$ because we are only interested in the failures due to cause 1 (abdominal surgery). We can estimate these using pseudo-observations as follows. Bayes' rule gives us 

\[
P(B_i > c | T_i \leq t, \delta_i = 1) = \frac{P(T_i \leq t, \delta_i = 1 | B_i > c) P(B_i > c)}{P(T_i \leq t, \delta_i = 1)},
\]
which is the ratio of a conditional cumulative incidence to the marginal cumulative incidence. This suggests the following estimators: 
\[
\widehat{TP}(t, c) = \frac{\sum_{i} \hat{C}^i_1(t) I(B_i > c)}{\sum_{i} \hat{C}^i_1(t)}
\]
and 
\[
\widehat{FP}(t, c) = \frac{\sum_{i} (1 - \sum_{j=1}^3\hat{C}^i_j(t)) I(B_i > c)}{\sum_{i} 1 - \sum_{j=1}^3\hat{C}^i_j(t)}.
\]

We conjecture that these estimators are asymptotically unbiased, based on the properties of the pseudo-observations shown in \citet{overgaard2017asymptotic} and application of the continuous mapping theorem. 

The ROC curve at time $t$ is a plot of the pairs $\{FP(t, c), TP(t, c)\}$ as the cutoff $c$ varies. The estimated area under the time varying ROC curve at a fixed time $t$ for signature $B$ is computed numerically using the trapezoidal rule, and is denoted $\widehat{AUC}(B, t)$. To our knowledge, this is a novel estimator of the time-varying ROC curve and, as a result, the time-varying AUC. The time-varying AUC has recently been shown to be a proper scoring function in the context of risk prediction of events at a fixed time, unlike the survival concordance index \citep{blanche2018c}, making it a sensible target for prediction. The Brier score is a proper scoring rule also, but is difficult to interpret in this case because the pseudo-observations are unbounded and because the true binary outcome is not completely observed. The time-varying predictiveness curve is then used to comprehensively evaluate the performance of the resulting prediction models.

We can also estimate a time varying version of the predictiveness curve \citep{pepe2007integrating} using pseudo-observations. The cause-specific predictiveness curve can be defined $P(T_i \leq t, \delta_i = 1 | B_i = c).$ We assess the usefulness of the model that creates $B_i$ by plotting the percentiles of $B_i$ versus the estimated cause specific cumulative incidence at those percentiles. As described by \citet{pepe2007integrating}, this curve displays whether there are meaningful variations in the risk of the event in question over the range of predicted quantities. To our knowledge, pseudo-observations have not been used to estimate predictiveness curves previously. 

This curve could be estimated by binning $B_i$, into deciles and taking the average of the pseudo-observations within each of the bins. Further, we can fit a locally weighted smooth average with the pseudo-observations as the outcome and the quantiles of $B_i$ as the predictor. The fitted values from this model provide an estimate of the predictiveness curve, from which we can display the estimated population cumulative incidence values as a smooth function of the predicted probabilities. 

The estimated true positive fractions and false positive fractions plotted as functions of the predicted risk values, combined with ROC and predictiveness curves provide a comprehensive summary of the performance and operating characteristics of the predictive models. We illustrate this in the data analysis below in Figure \ref{predfig}. 

\subsection{Super-Learner predictive model development}

It is unknown which class of prediction procedures will perform best on a given real dataset. Instead of selecting one specific procedure or machine learning method, we use an ensemble of methods, and then form predictions by stacking the predictions from the particular methods into an "ensemble predictor". The stacking is performed by selecting the linear combination of individual predictions that maximizes the estimated $AUC(t)$ in a cross-validation procedure. This stacked modeling approach is called the Superlearner \citep{van2007super}, and is implemented in the \texttt{SuperLearner} R package \citep{superlearner}. \citet{van2007super} demonstrate that the resulting prediction algorithm performs at least as well as the best predictor in the library of algorithms. The general algorithm is described in that paper, and here we present the specific steps adapted to our setting.

Let $\mathcal{L}$ denote the library of $K$ prediction algorithms. In general, a prediction algorithm with index $k$ for $k = 1, \ldots, K$, could be any method to develop $\hat{f}_k$. The specific algorithms that we use in our simulations and data example are given in Table \ref{library}. The steps are as follows: 

\begin{enumerate}
\item For a fixed, finite grid of time points $\{t_1, \ldots, t_{\star}, \ldots, t_m\}$, where $t_{\star}$ is the time point of interest for predicting incidence, calculate the pseudo-observations $\hat{C}^i_j(t_l)$, for $j = 1, 2, 3; l = 1, \ldots, m$; $i = 1, \ldots, n$. 
\item Split the dataset $\mathcal{X}$ into a training and validation sample, according to a V-fold cross validation scheme: split the ordered $n$ observations into $V$ equal size groups. Let the $\nu$-th group be the validation sample, and the remaining group the training sample, $\nu = 1,...,V$. Define $T(\nu)$ to be the $\nu$th training data split and $V(\nu)$ to be the corresponding validation data split where $T(\nu) = \mathcal{X} \setminus V (\nu), \nu = 1,...,V$.

\item Stack the datasets so that there are $m$ rows per subject $i$, with the same covariates $\mathbf{X}_i$, additional variables for the pseudo-observations, and an additional variable for the times $t_1, \ldots, t_{\star}, \ldots, t_m$. 
\item Fit each algorithm on the training set for each fold $T(\nu)$, to obtain $\hat{f}_{k\nu}$ for $k = 1, \ldots, K$ and $\nu = 1,...,V$. 
\item For each $\hat{f}_{k\nu}$, obtain predicted pseudo-observations at $t_{\star}$ for the validation set for each fold, i.e., $\tilde{C}^i_1(t)_{k} = \hat{f}_{k\nu}(X_i, t_{\star})$, for $i \in V(\nu)$ and $k = 1, \ldots, K$. This yields $K$ predicted pseudo-observations for each subject $i$. Denote the vector of these for subject $i$ and time $t_{\star}$ as $\mathbf{\tilde{C}}^i_1(t_{\star})$.
\item For a fixed $\lambda$ and $t_{\star}$, find the $K$ dimensional parameter vector $\hat{\alpha}$ that satisfies 
		\[
        \max_\alpha \widehat{AUC}(\alpha^T \mathbf{\tilde{C}}^i_1(t_{\star}), t_{\star}) + \lambda \sum_k |\alpha_k|. 
        \]
        To ensure proper calibration, normalize the coefficients by their sum, $\hat{\alpha}^* = \hat{\alpha} / \mathbf{1}^T \hat{\alpha}$. Although multiple time points may be used in the prediction procedures, note that only the time point $t_{\star}$ is used in the loss function optimization procedure.  
\item For $k = 1, \dots, K$, fit each algorithm on the full dataset to obtain $\hat{f}_k(\mathbf{X}_i, t_{\star})$, for $i = 1, \dots, n$. This forms an $n$ by $k$ matrix where the rows are predicted pseudo-observations for each subject, and the columns correspond to different prediction methods. The final predictor is the linear combination $\hat{\alpha}^{*T}\hat{f}_k$ according to the cross-validated weights obtained in the previous step.
\end{enumerate}

The penalty term in step 6 above is introduced to solve the indentifiability issues that arise in maximization of the AUC, similar to what was suggested by \citet{lin2011selection} and further discussed by \citet{fong2016combining}. Briefly, the issues in maximizing the empirical AUC arise because the ROC curve is invariant to monotone transformations of the signature. Penalization of the objective function is one way to solve this issue and yield a unique solution for $\alpha$. 

The calculation of the pseudo-observations at the fixed grid of time points in step (a) and the stacking thereof allows us to borrow information across the time points, with the idea that this should perform as well or better than the counting process approach, or using an single time point. The computational burden of both calculating the pseudo-observations and fitting the models in the library increases with the number of time points. The selection of the density and location of the time points in the grid is not something we investigate in this paper, however, in principle, one can include all observed event times. We suggest possible approaches to this issue in the discussion.

\section{Simulation study}

\subsection{Data generation}
For each replicate of the simulation scenario, we generate a dataset with 500 subjects, 20 normally distributed variables that are correlated with each other to varying degrees, and event times that are distributed Weibull with about 20\% cumulative incidence of the surgery event and 7\% cumulative incidence of the competing event of death at time $t_{\star}$. Censoring is completely independent (except for scenario D) and is either 20\% or 50\% on average random censoring, corresponding to 33\% or 65\% at the time $t_{\star}$. In scenario D, we allow censoring to depend on covariates. In all scenarios, the competing event was generated according to a Weibull distribution with a scale parameter that depends on a single covariate that was not related to the covariates that are associated with the time to surgery. 

For each simulated dataset, we implement our proposed method by computing pseudo observations for a grid of 4 time points, 2 before and 1 after the time point of interest, $t_\star$ and using the AUC based SuperLearner procedure desribed in detail above. For comparison to our proposed method, we fit a pseudo-observation model using a single time point at $t_{\star}$, a IPCW SuperLearner model on the binary outcome created by dichotomizing the time to surgery at $t_{\star}$ using the non-negative log likelihood loss, the model proposed by \citet{binder2009boosting} as implemented in the \texttt{CoxBoost} R package, and the random forests for competing risks survival proposed by  \citep{ishwaran2014random}, as implemented in the R package \texttt{randomForestsSRC}. We estimate and use inverse probability of censoring weighting in the binary SuperLearner approach via Kaplan-Meier for censoring event. For the binary SuperLearner only methods that natively allowed for weights were considered, while more libraries were used by the pseudo-observations. Table \ref{library} in the application section describes the libraries used for each SuperLearner type. We generate an independent dataset on which to estimate the out-of-sample prediction error as estimated by the AUC. 

We consider 2 different censoring rates and 4 different scenarios:

\begin{itemize} 
\item Scenario A: exactly one variable is associated with the scale parameter of the surgery outcome in a log-linear way. 
\item Scenario B: exactly 5 variables are associated with the scale parameter of the surgery outcome according to a b-spline relationship of degree 3, and with a linear interaction effect.
\item Scenario C: exactly 5 variables are associated with the scale parameter of the surgery outcome in a complex way, including non-linearities, interactions, and threshold effects. 
\item Scenario D: Same as B, but censoring depends on a subset of the covariates from those associated with the surgery outcome. The covariate effect on censoring is log-linear on the scale parameter of a Weibull distribution. 
\item Scenario E: Bootstrap of 500 samples and the first 20 covariates from the CD data analysis. In this case the true event times and probabilities are unknown. 
\end{itemize}

Detailed specification of and code for the simulation scenarios is given in the supplementary materials. In addition, we released an R package called \texttt{sachsmc/pseupersims} on github that contains the code used to perform this simulation study. 

The purpose of the simulation study is to evaluate the ability of the SuperLearner model based on pseudo-observations to develop a prediction model for the cumulative incidence and to compare to alternative approaches, and to the true model. 

\subsection{Using pseudo-observations to develop prediction models}  

Our proposed approach performs comparably to competing methods in terms of accurately predicting the true cumulative incidence, as shown in Tables \ref{sim1} and \ref{sim2}. The average AUC over the simulation replicates is superior to all other considered approaches in scenarios B, C, and D, while the other approaches perform well in scenario A. Compared to the binary SuperLearner approach, differences in the performance of the pseudo-observation based models were small, while the CoxBoost model and random forests occasionally showed markedly inferior performance. Still the classification accuracy of the pseudo observation model is competitive with other approaches and potential for performance gains are clear. In the simulation E, 500 person bootstraps from the CD analysis data, we present the average estimated AUC for each of the methods. In that case, even the model based pseudo observations estimated at a single time point outperforms the binary, but the CoxBoost and random forests models outperform all other methods. 

 The models based on pseudo observations computed at a single time point occasionally had convergence issues and therefore have extreme outliers, which explains the sometimes dramatically high MSE. The pseudo-observation model based on four time points had smaller variance in the predictions relative to the binary model in scenarios D and E. However, in some cases such as scenario B, we observe a higher variance, because, unlike all of the other methods, the predictions from the pseudo-observation models are not bounded. Finally, our proposed estimator for the AUC using the pseudo observations appears to be unbiased, as the estimated values were nearly identical to the AUC estimated using the true, and typically unobserved binary outcome of surgery at time $t_{\star}$. 

\begin{footnotesize}
\begin{table}
\caption{\label{sim1}Results of simulation study using a setting with approximately 20\% random censoring (approximately 33\% missingness in the binary outcome). We compare the estimation of the area under the time varying ROC curve using the true binary outcome that is unobserved (labelled tbauc), the estimation of the AUC using pseudo observations as described in Section 2 (labelled pauc), and the bias, standard deviation, and mean squared error (mse) of the predicted outcomes compared to the true probabilities of the outcome. }
\centering
\begin{tabular}{lrrrrrrr}
\hline
model & mean.tbauc & sd.tbauc & mean.pauc & sd.pauc & mean.bias & sd.prob & mse\\
\hline
& \multicolumn{7}{c}{Scenario A} \\
\hline
binary & 0.735 & 0.031 & 0.736 & 0.033 & 0.003 & 0.146 & 0.043\\
CoxBoost & 0.734 & 0.026 & 0.731 & 0.029 & 0.002 & 0.116 & 0.040\\
pseudo & 0.729 & 0.028 & 0.729 & 0.031 & 0.002 & 0.147 & 0.086\\
pseudo.single & 0.717 & 0.033 & 0.718 & 0.035 & 0.000 & 0.171 & 0.132\\
rfsrc & 0.716 & 0.027 & 0.717 & 0.029 & -0.008 & 0.123 & 0.049\\
true & 0.747 & 0.025 & 0.747 & 0.028 & 0.000 & 0.228 & 0.000\\
\hline
& \multicolumn{7}{c}{Scenario B} \\
\hline
binary & 0.569 & 0.042 & 0.570 & 0.045 & -0.001 & 0.064 & 0.068\\
CoxBoost & 0.522 & 0.034 & 0.514 & 0.038 & 0.009 & 0.041 & 0.044\\
pseudo & 0.546 & 0.073 & 0.548 & 0.075 & -0.056 & 0.505 & 103.615\\
pseudo.single & 0.533 & 0.051 & 0.535 & 0.054 & -0.009 & 0.527 & 5.799\\
rfsrc & 0.579 & 0.036 & 0.580 & 0.040 & -0.011 & 0.072 & 0.055\\
true & 0.666 & 0.031 & 0.665 & 0.035 & 0.000 & 0.122 & 0.000\\
\hline
& \multicolumn{7}{c}{Scenario C} \\
\hline
binary & 0.839 & 0.039 & 0.840 & 0.041 & 0.010 & 0.256 & 0.062\\
CoxBoost & 0.630 & 0.032 & 0.622 & 0.036 & 0.002 & 0.081 & 0.025\\
pseudo & 0.848 & 0.062 & 0.848 & 0.062 & 0.010 & 0.268 & 0.257\\
pseudo.single & 0.833 & 0.040 & 0.834 & 0.042 & -0.007 & 0.300 & 0.117\\
rfsrc & 0.780 & 0.034 & 0.780 & 0.036 & -0.010 & 0.109 & 0.034\\
true & 0.927 & 0.013 & 0.927 & 0.013 & 0.000 & 0.521 & 0.000\\
\hline
& \multicolumn{7}{c}{Scenario D} \\
\hline
binary & 0.839 & 0.037 & 0.843 & 0.037 & 0.009 & 0.258 & 0.054\\
CoxBoost & 0.625 & 0.032 & 0.621 & 0.034 & -0.002 & 0.081 & 0.027\\
pseudo & 0.853 & 0.035 & 0.858 & 0.036 & 0.011 & 0.252 & 0.155\\
pseudo.single & 0.830 & 0.075 & 0.834 & 0.075 & 0.010 & 0.380 & 4.797\\
rfsrc & 0.782 & 0.032 & 0.785 & 0.033 & -0.013 & 0.107 & 0.042\\
true & 0.926 & 0.013 & 0.929 & 0.014 & 0.000 & 0.520 & 0.000\\
\hline
& \multicolumn{7}{c}{Scenario E: bootstrap of CD data, 36\% missingness} \\
\hline
binary & - & - & 0.525 & 0.035 & - & 0.324 & -\\
CoxBoost & - & - & 0.585 & 0.043 & - & 0.062 & -\\
pseudo & - & - & 0.560 & 0.059 & - & 0.297 & -\\
pseudo.single & - & - & 0.552 & 0.061 & - & 0.184 & -\\
rfsrc & - & - & 0.598 & 0.042 & - & 0.089 & -\\
\hline
\end{tabular}
\end{table}
\end{footnotesize}

\begin{footnotesize}
\begin{table}
\caption{\label{sim2}Results of simulation study using a setting with approximately 50\% random censoring (approximately 65\% missingness in the binary outcome). We compare the estimation of the area under the time varying ROC curve using the true binary outcome that is unobserved (labeled tbauc), the estimation of the AUC using pseudo observations as described in Section 2 (labeled pauc), and the bias, standard deviation, and mean squared error (mse) of the predicted outcomes compared to the true probabilities of the outcome. }
\centering
\begin{tabular}{lrrrrrrr}
\hline
model & mean.tbauc & sd.tbauc & mean.pauc & sd.pauc & mean.bias & sd.pred & $100 * $mse\\
\hline
& \multicolumn{7}{c}{Scenario A} \\
\hline
binary & 0.730 & 0.034 & 0.728 & 0.039 & 0.024 & 0.140 & 0.114\\
CoxBoost & 0.734 & 0.028 & 0.729 & 0.036 & 0.026 & 0.110 & 0.117\\
pseudo & 0.722 & 0.041 & 0.719 & 0.048 & 0.037 & 0.188 & 8.272\\
pseudo.single & 0.704 & 0.041 & 0.701 & 0.045 & 0.031 & 0.161 & 0.168\\
rfsrc & 0.711 & 0.031 & 0.708 & 0.038 & 0.017 & 0.110 & 0.077\\
true & 0.747 & 0.028 & 0.745 & 0.035 & 0.000 & 0.230 & 0.000\\
\hline
& \multicolumn{7}{c}{Scenario B} \\
\hline
binary & 0.552 & 0.043 & 0.554 & 0.051 & 0.020 & 0.061 & 0.123\\
CoxBoost & 0.519 & 0.033 & 0.513 & 0.042 & 0.032 & 0.042 & 0.144\\
pseudo & 0.532 & 0.065 & 0.534 & 0.069 & 0.013 & 0.382 & 6.204\\
pseudo.single & 0.517 & 0.044 & 0.517 & 0.051 & 0.122 & 0.708 & 96.630\\
rfsrc & 0.557 & 0.035 & 0.558 & 0.045 & 0.013 & 0.067 & 0.068\\
true & 0.661 & 0.029 & 0.659 & 0.037 & 0.000 & 0.121 & 0.000\\
\hline
& \multicolumn{7}{c}{Scenario C} \\
\hline
binary & 0.814 & 0.039 & 0.802 & 0.043 & 0.033 & 0.231 & 0.182\\
CoxBoost & 0.620 & 0.034 & 0.607 & 0.044 & 0.044 & 0.073 & 0.232\\
pseudo & 0.820 & 0.057 & 0.811 & 0.057 & 0.075 & 0.319 & 15.444\\
pseudo.single & 0.792 & 0.074 & 0.780 & 0.075 & 0.045 & 0.320 & 3.285\\
rfsrc & 0.738 & 0.036 & 0.726 & 0.046 & 0.035 & 0.089 & 0.161\\
true & 0.926 & 0.013 & 0.911 & 0.017 & 0.000 & 0.521 & 0.000\\
\hline
& \multicolumn{7}{c}{Scenario D} \\
\hline
binary & 0.822 & 0.042 & 0.812 & 0.047 & 0.026 & 0.234 & 0.139\\
CoxBoost & 0.616 & 0.033 & 0.602 & 0.040 & 0.031 & 0.075 & 0.139\\
pseudo & 0.824 & 0.040 & 0.814 & 0.043 & 0.032 & 0.322 & 0.403\\
pseudo.single & 0.788 & 0.086 & 0.781 & 0.087 & 0.016 & 0.432 & 6.134\\
rfsrc & 0.738 & 0.035 & 0.727 & 0.039 & 0.023 & 0.090 & 0.090\\
true & 0.925 & 0.013 & 0.911 & 0.017 & 0.000 & 0.522 & 0.000\\
\hline
\end{tabular}
\end{table}
\end{footnotesize}

\section{CD complications results}

The total sample size included 9605 subjects diagnosed with adult-onset CD after January 1, 2003. The demographic characteristics we considered for prediction include gender, age at diagnosis, calendar year of diagnosis, highest parental education, whether or not parents were born outside of Sweden, whether the subject was born outside of Sweden, and whether there is any family history of CD or ulcerative colitis. Clinical characteristics considered at the time of diagnosis are whether it was an inpatient or outpatient diagnosis, Montréal classification of disease, presence of primary sclerosing cholangitis, presence of extra-intestinal manifestations, and exposure to CD related medications. Comorbidities considered were diabetes, hypertension, ischemic heart disease, cerebrovascular disease, congestive heart failure, chronic obstructive pulmonary disorder, and kidney failure. The covariates are described in Supplemental Tables 1 - 3.

\begin{figure}[ht]
\centering
\includegraphics[scale = .7]{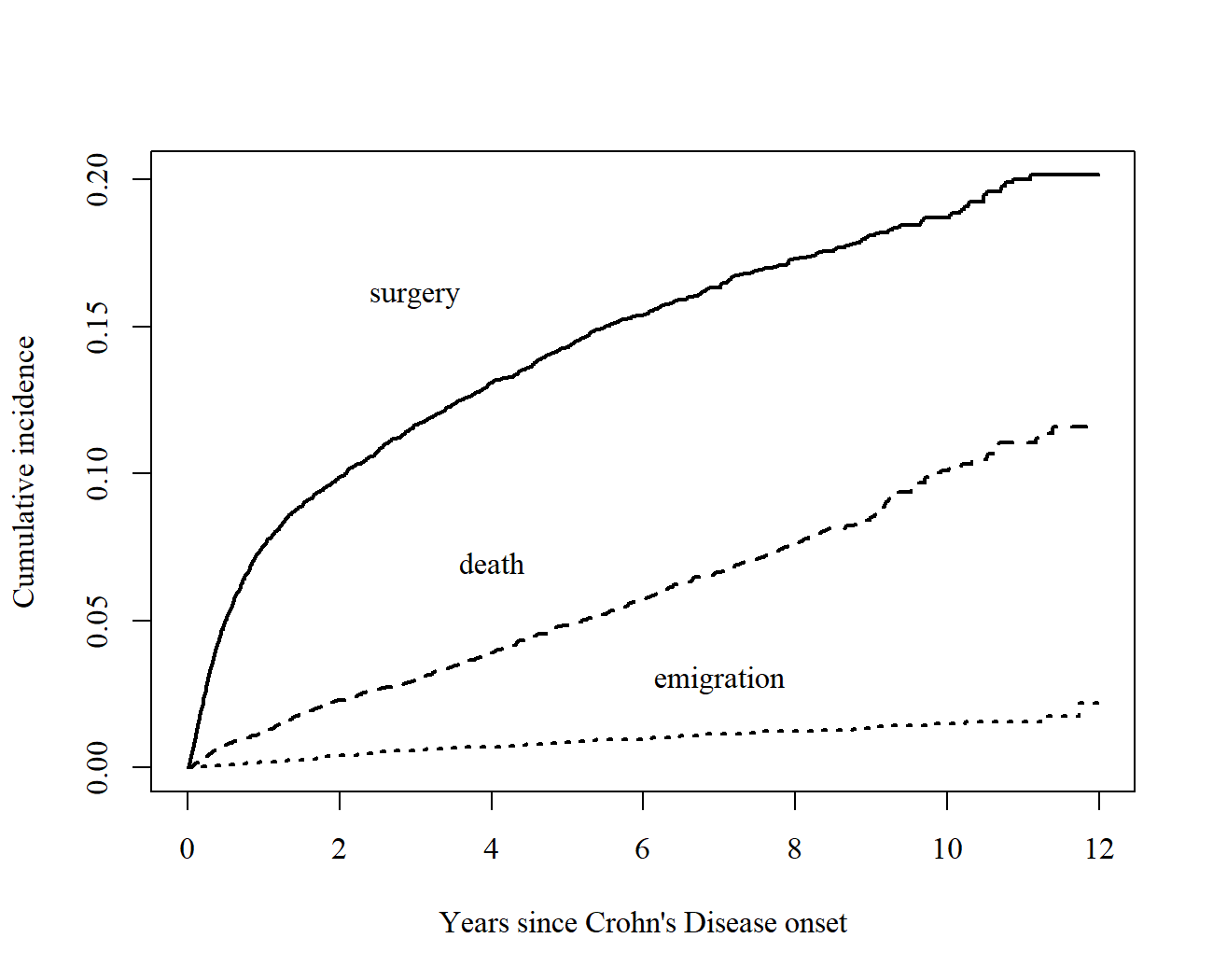}
\caption{Cumulative incidence curves of surgery, death or emigration since onset of Crohn's disease in a sample of 9605 patients with onset between 2003 and 2014. \label{cifig}}
\end{figure}

The cumulative incidence curves for the two competing risks are shown in Figure \ref{cifig}. We are interested in developing a model to predict the incidence of abdominal surgery within 5 years, while accounting for the competing risks of death and emigration. 

\subsection{Predictive performance}

We randomly split the dataset into two, equal-size subsamples, a training and a validation set. On the training set, we performed the algorithm as described above, including cross-validation within the training sample. Time of entry into the study (CD diagnosis) was categorized into the sets January 2003-December 2008, and then in 6 month periods from January 2009 to December 2015. Pseudo observations were calculated conditionally on the categorized time of entry to account for censoring that is determined by time of entry and for subsets with no subjects with more than 5 years of follow up, the last estimated value of the cumulative incidence was carried forward. We used the grid of time points $\{2.5, 4, 5, 5.5\}$ years, and calculated pseudo observations at these times. We ran one SuperLearner model using the full grid, and another one including all nested subsets of the grid (meaning no non-adjacent times were included) combined with each ML procedure in the ensemble. For comparison, we ran all methods considered in the simulations. The binary outcome had 36\% missingness on the training sample. The censoring weights used in the binary method were estimated using Kaplan-Meier for censoring, stratified by the times of entry in the same way as for the pseudo observations. 

The performance of the candidate learning algorithms are described in Table \ref{library} by the cross-validated AUCs and the estimated SuperLearner coefficients, when applicable. Precise specification of the various settings and tuning parameters for each procedure are supplied in the supplementary materials. The performance of each of the models is modest, with AUCs ranging from 0.57 to 0.64. The non-negative log-likelihood risk is also presented in Table \ref{library} for the binary methods; smaller values of this indicate better performance. The weights estimated using the SuperLearner ensemble optimization are given in the weight columns. The performance of the ensemble over all nested subsets of the times in the grid was similar, but slightly superior, to that using only the four point time grid. 

\begin{figure}[ht]
\centering
\includegraphics[width=.95\textwidth]{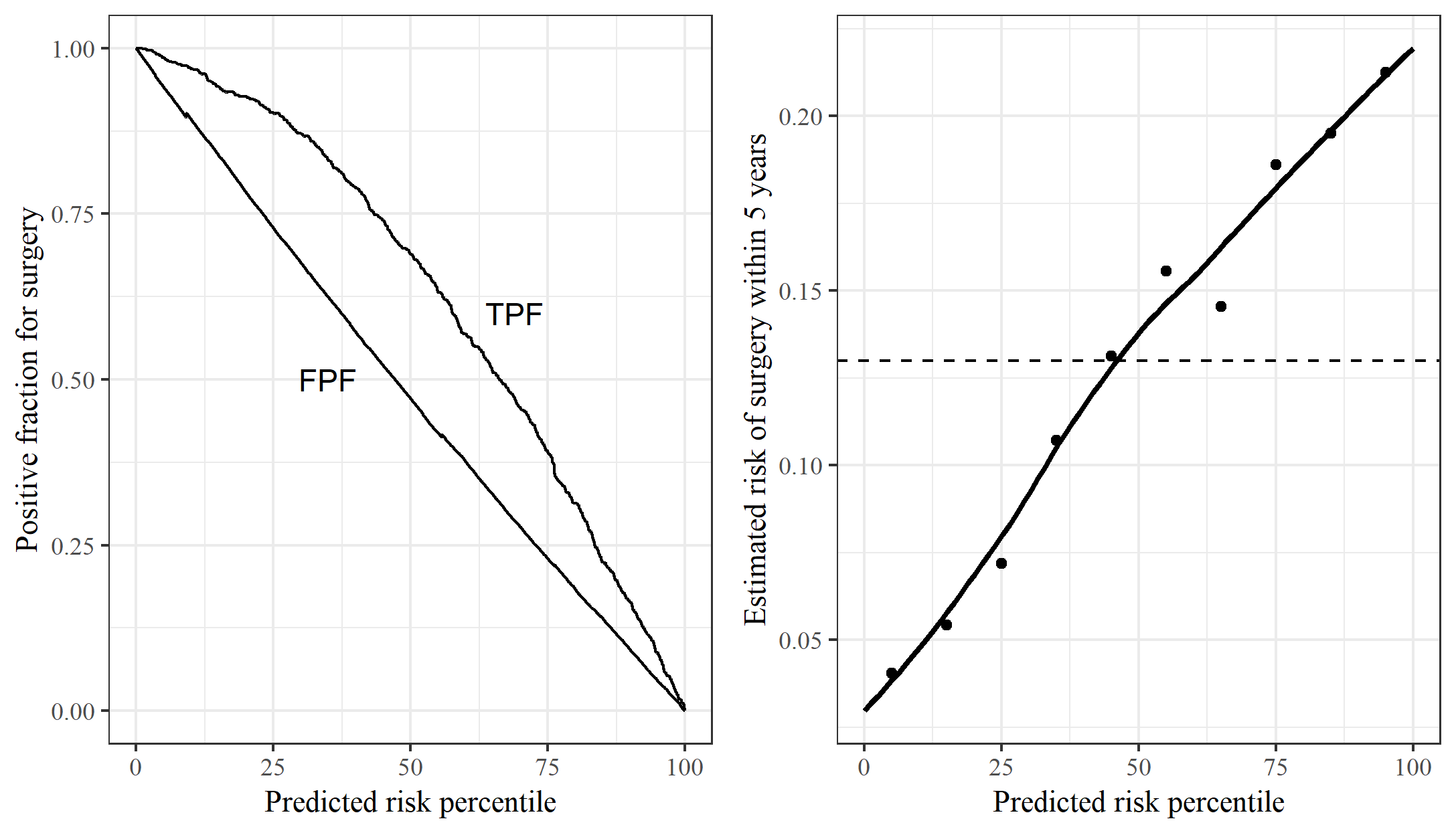}
\caption{(Left panel) Estimated out-of-sample true and false positive fractions as functions of the risk percentiles for the cumulative incidence of surgery at 5 years, accounting for the competing risk of death, since onset of Crohn's disease in the validation sample. (Right panel) Estimated predictiveness curves for the cumulative incidence of surgery at 5 years, accounting for the competing risk of death, since onset of Crohn's disease in the validation sample. Points are binned estimates, solid line is the locally weighted smoothed average of the individual pseudo-observations (not the binned estimates), and dotted line is the marginal risk. \label{predfig}}
\end{figure}

The out-of-sample predictive performance results are shown in Figure \ref{predfig}. The SuperLearner model using the pseudo-observations was superior to the weighted binary SuperLearner, with estimated AUCs of 0.64 and 0.50, respectively. The random forests model and the CoxBoost model performed slightly worse than the pseudo-observation approach, as summarized by the AUC, with an AUC of 0.632 and 0.60, respectively. The characteristics of the models differed in other ways, such as the shape of the ROC curve and predictiveness curves. 

The predictiveness curves in Figure \ref{predfig} show that the pseudo-observation model is informative regarding risk of abdominal surgery. CD patients in the top percentile of the predicted risk have an estimated 21\% probability of surgery within 5 years, compared to the marginal risk of 13\%. Likewise, the patients with the lowest predicted risk have less than a 5\% risk of surgery within 5 years. The smoothed locally weighted average of the pseudo observations allows us to estimate the predictiveness curve as a smooth function of the risk percentile, as shown in the solid line. This also suggests that pseudo-observation based predictions have value, as it varies monotonically over the percentiles of the predictions.

\begin{small}
\begin{table}
\caption{\label{library} Description of the prediction algorithms used and in which settings they were used for the example. For the binary and pseudo observations SuperLearner approaches, we summarize the non-negative log-likelihood risk for binary (smaller is better), and the AUC (larger is better) for pseudo for each method, along with the estimated coefficients for the method in the ensemble. The nested pseudo column presents the average (over the subsets) AUC and sum of the weights for each model in which all nested subsets of the grid of time points are allowed in the ensemble. For the survival based methods, CoxBoost and random forests, we compute the out-of-sample AUC based on pseudo-observations for comparison to the other methods. }
\fbox{
\begin{tabular}{p{4.4cm}|c|rr|rr|rr}
\textbf{Model}  & \textbf{R package}  & \multicolumn{2}{|c|}{Binary} & \multicolumn{2}{|c}{All Pseudo} &
\multicolumn{2}{|c}{Nested Pseudo} \\
& & Risk & weight & AUC & weight & AUC & weight\\
\hline
\multicolumn{6}{c}{Used for both Binary and Pseudo Observations} \\
\hline
General Linear model, with screening & stats  & 25.2 & 0.17 & 0.63 & 0.15 & 0.63 & 0.04\\
\hline
Generalized additive model, with screening & gam  & 24.7 & 0.00 &0.63 &0.19 & 0.63& 0.03\\
\hline
Random forest with bagging & randomForest  & 22.1 & 0.40& 0.59 &0.20& 0.60 & 0.04\\
\hline
Recursive partitioning and regression trees & rpart & Inf  &0.08& 0.57 &0.04& 0.55& 0.14\\
\hline
LASSO & glmnet  & 22.6 & 0.00 & 0.64 & 0.20& 0.63 & 0.03\\
\hline
Multivariate adaptive polynomial spline, with screening & polspline  &27.6 &0.06 & 0.61 & 0.10&0.62 & 0.21\\
\hline
Extreme gradient boosting & xgboost & 22.0 & 0.28 & 0.63 & 0.46& 0.63& 0.25\\
\hline
\multicolumn{6}{c}{Used only for Pseudo Observations}\\
\hline
Kernel support vector machine & kernlab  & &&0.59&$-0.07$ & 0.58 & 0.08\\
\hline
Stepwise selection & stats & &&0.63&0.13& 0.63 & 0.05 \\
\hline
\hline
\multicolumn{6}{c}{Used only for censored survival outcomes}\\
\hline
Cox Boost & CoxBoost & &&0.63&&&\\
\hline
Random Forests & randomForestSRC & & &0.60&&&\\
\end{tabular}
}
\end{table}
\end{small}

\begin{figure}[ht]
\centering
\includegraphics{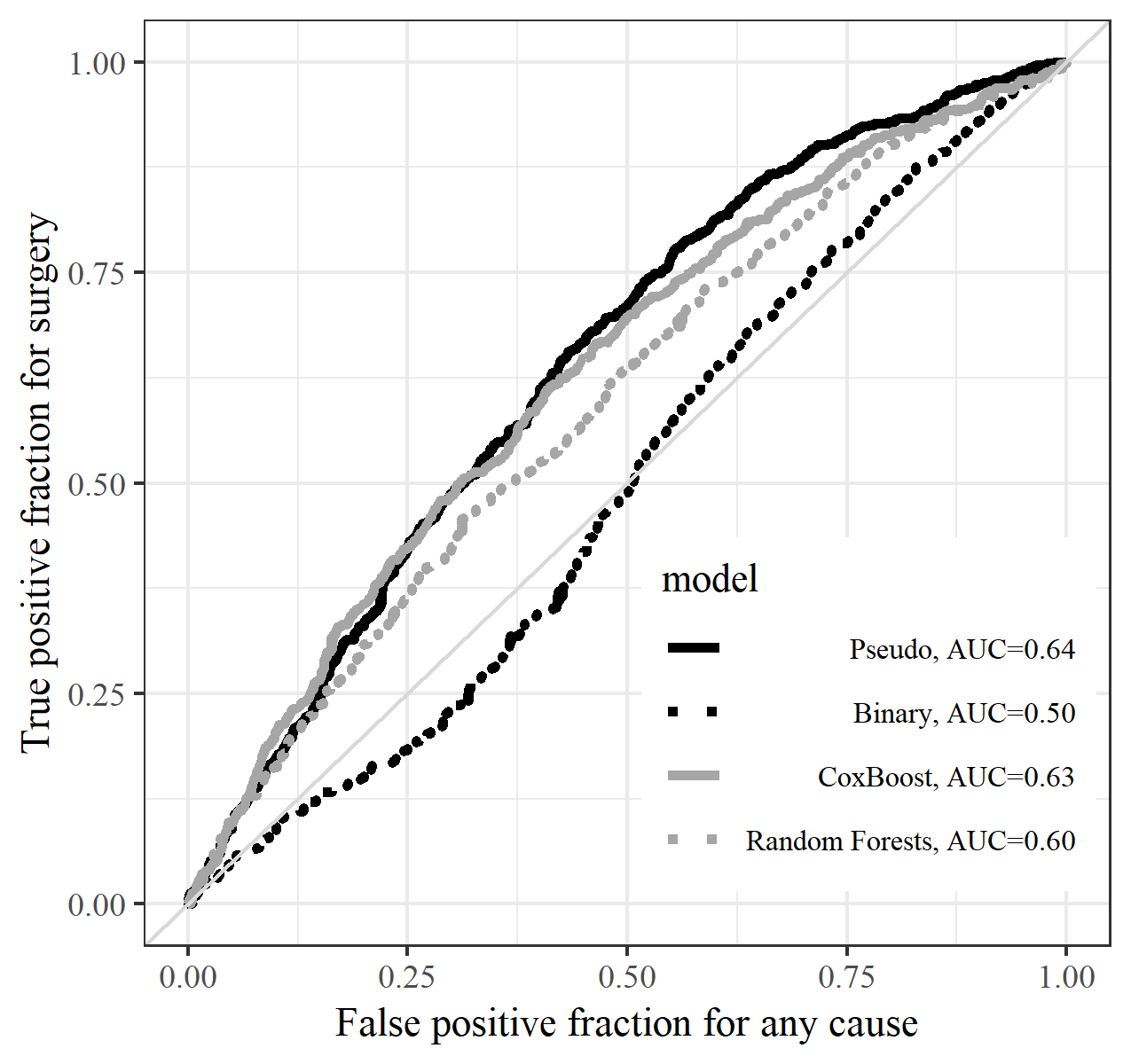}
\caption{Estimated out of sample ROC curves for the different models used in the example. \label{compfig}}
\end{figure}

\section{Discussion}
In our motivating study, we developed a predictive model for abdominal surgery within 5 years following CD diagnosis based on routinely collected clinical characteristics observed at the time of diagnosis. The proposed prediction procedure has an estimated out of sample AUC of 0.64, which is the highest among the considered methods. Medical management of CD is becoming increasingly personalized and individualized risk estimations are needed. For some patients at high risk of surgery, a more aggressive medication or monitoring strategy may be warranted. In some patients, where a surgical intervention is inevitable, it should be performed sooner to improve quality of life. 

Although it may be helpful to further understanding, it is not necessary to 'unpack the black box' for the predictive model to be useful in this setting because all variables contributing to the predictions will be collected regardless of their usefulness in predicting risk. For this reason, the model for the predictions could be used directly, via electronic health records, for real-time predictions in clinical settings, without the need to know which variables are driving the models. Implementation of such real-time predictions and evaluating its impact on patient outcomes is a future area of clinical research for the authors. 

We have shown that the use of pseudo-observations based on the cumulative incidence can be used with the all libraries for continuous data that are compatible with the SuperLearner package, and by extension can be used with any existing or newly developed ML procedures for continuous data in the same way. We also show that our proposed loss function, the pseudo-observation based time-dependent AUC, is a useful measure of predictive value for our setting and that the pseudo-observation based predictiveness curve is a useful depiction of those predictions. We provide freely available software which contains this modification to the SuperLearner package in addition to the other methods described herein.

Although our proposed pseudo-observation based method outperforms the IPCW binary method in terms of predictive accuracy in the data analysis and when using multiple times points in the most settings in the simulations, it performs more similarly when a single time point is used. In addition, in most, but not all, of the scenarios, our method outperforms the individual class survival methods. The pseudo-observation based method, however, we found easier to apply, as the package \texttt{randomForestSRC} caused R to crash repeatedly during development; we believe this may be due to the large sample size leading to memory issues in the data analysis. Although our proposed methods performs similarly to, sometimes better and sometimes worse than, the other existing methods, a new ML procedure may be developed that will improve performance of our methods in the future. As creating pseudo-observations is a general pre-processing step, a pseudo-observation based ensemble could include this new method without extension, where direct adaptations of the method for application to censored data will take time and effort. 

The idea of stacking pseudo-observations in regression models is not new, and was an approach suggested for dynamic prediction models of competing risks \citep{nicolaie2013dynamic}. One advantage of the SuperLearner approach is that it is possible to include multiple time-point sets, as we did in the data analysis, to determine what time points from a pre-selected grid are most useful. The question of how to choose the grid of time points at which to calculate the pseudo-observations is one that we did not investigate in great detail. It is of note that we selected the grid somewhat arbitrarily, thus one might expect that performance can only be improved via some selection or optimization process. The pre-selection of the grid is an avenue for future research.

\section{Bibliography}

\bibliographystyle{rss}
\bibliography{irisbiblio}

\begin{thebibliography}{30}
\expandafter\ifx\csname natexlab\endcsname\relax\def\natexlab#1{#1}\fi
\expandafter\ifx\csname url\endcsname\relax
  \def\url#1{\texttt{#1}}\fi
\expandafter\ifx\csname urlprefix\endcsname\relax\def\urlprefix{URL: }\fi

\bibitem[{Aalen and Johansen(1978)}]{aalen1978empirical}
Aalen, O.~O. and Johansen, S. (1978) An empirical transition matrix for
  non-homogeneous markov chains based on censored observations.
\newblock \textit{Scandinavian Journal of Statistics}, 141--150.

\bibitem[{Andersen et~al.(2003)Andersen, Klein and
  Rosth{\o}j}]{andersen2003generalised}
Andersen, P.~K., Klein, J.~P. and Rosth{\o}j, S. (2003) Generalised linear
  models for correlated pseudo-observations, with applications to multi-state
  models.
\newblock \textit{Biometrika}, \textbf{90}, 15--27.

\bibitem[{Andersen and Pohar~Perme(2010)}]{andersen2010pseudo}
Andersen, P.~K. and Pohar~Perme, M. (2010) Pseudo-observations in survival
  analysis.
\newblock \textit{Statistical methods in medical research}, \textbf{19},
  71--99.

\bibitem[{Binder et~al.(2009)Binder, Allignol, Schumacher and
  Beyersmann}]{binder2009boosting}
Binder, H., Allignol, A., Schumacher, M. and Beyersmann, J. (2009) Boosting for
  high-dimensional time-to-event data with competing risks.
\newblock \textit{Bioinformatics}, \textbf{25}, 890--896.

\bibitem[{Binder et~al.(2014)Binder, Gerds and Andersen}]{binder2014pseudo}
Binder, N., Gerds, T.~A. and Andersen, P.~K. (2014) Pseudo-observations for
  competing risks with covariate dependent censoring.
\newblock \textit{Lifetime data analysis}, \textbf{20}, 303--315.

\bibitem[{Blanche et~al.(2018)Blanche, Kattan and Gerds}]{blanche2018c}
Blanche, P., Kattan, M.~W. and Gerds, T.~A. (2018) The c-index is not proper
  for the evaluation of $ t $-year predicted risks.
\newblock \textit{Biostatistics}.

\bibitem[{Cortese et~al.(2013)Cortese, Gerds and
  Andersen}]{cortese2013comparing}
Cortese, G., Gerds, T.~A. and Andersen, P.~K. (2013) Comparing predictions
  among competing risks models with time-dependent covariates.
\newblock \textit{Statistics in medicine}, \textbf{32}, 3089--3101.

\bibitem[{Fine and Gray(1999)}]{fine1999proportional}
Fine, J.~P. and Gray, R.~J. (1999) A proportional hazards model for the
  subdistribution of a competing risk.
\newblock \textit{Journal of the American statistical association},
  \textbf{94}, 496--509.

\bibitem[{Fong et~al.(2016)Fong, Yin and Huang}]{fong2016combining}
Fong, Y., Yin, S. and Huang, Y. (2016) Combining biomarkers linearly and
  nonlinearly for classification using the area under the roc curve.
\newblock \textit{Statistics in medicine}, \textbf{35}, 3792--3809.

\bibitem[{Gerds and Ozenne(2018)}]{gerds2018}
Gerds, T.~A. and Ozenne, B. (2018) \textit{riskRegression: Risk Regression
  Models and Prediction Scores for Survival Analysis with Competing Risks}.
\newblock \urlprefix\url{https://CRAN.R-project.org/package=riskRegression}.
\newblock R package version 2018.10.03.

\bibitem[{Gomoll{\'o}n et~al.(2016)Gomoll{\'o}n, Dignass, Annese, Tilg,
  Van~Assche, Lindsay, Peyrin-Biroulet, Cullen, Daperno, Kucharzik
  et~al.}]{gomollon20163rd}
Gomoll{\'o}n, F., Dignass, A., Annese, V., Tilg, H., Van~Assche, G., Lindsay,
  J.~O., Peyrin-Biroulet, L., Cullen, G.~J., Daperno, M., Kucharzik, T. et~al.
  (2016) 3rd european evidence-based consensus on the diagnosis and management
  of crohn’s disease 2016: part 1: diagnosis and medical management.
\newblock \textit{Journal of Crohn's and Colitis}, \textbf{11}, 3--25.

\bibitem[{Graw et~al.(2009)Graw, Gerds and Schumacher}]{graw2009pseudo}
Graw, F., Gerds, T.~A. and Schumacher, M. (2009) On pseudo-values for
  regression analysis in competing risks models.
\newblock \textit{Lifetime Data Analysis}, \textbf{15}, 241--255.

\bibitem[{Heagerty et~al.(2000)Heagerty, Lumley and Pepe}]{heagerty2000time}
Heagerty, P.~J., Lumley, T. and Pepe, M.~S. (2000) Time-dependent roc curves
  for censored survival data and a diagnostic marker.
\newblock \textit{Biometrics}, \textbf{56}, 337--344.

\bibitem[{Hothorn et~al.(2006)Hothorn, B{\"u}hlmann, Dudoit, Molinaro and Van
  Der~Laan}]{hothorn2005survival}
Hothorn, T., B{\"u}hlmann, P., Dudoit, S., Molinaro, A. and Van Der~Laan, M.~J.
  (2006) Survival ensembles.
\newblock \textit{Biostatistics}, \textbf{7}, 355--373.

\bibitem[{Ishwaran et~al.(2014)Ishwaran, Gerds, Kogalur, Moore, Gange and
  Lau}]{ishwaran2014random}
Ishwaran, H., Gerds, T.~A., Kogalur, U.~B., Moore, R.~D., Gange, S.~J. and Lau,
  B.~M. (2014) Random survival forests for competing risks.
\newblock \textit{Biostatistics}, \textbf{15}, 757--773.

\bibitem[{Jacobsen and Martinussen(2016)}]{jacobsen2016note}
Jacobsen, M. and Martinussen, T. (2016) A note on the large sample properties
  of estimators based on generalized linear models for correlated
  pseudo-observations.
\newblock \textit{Scandinavian Journal of Statistics}, \textbf{43}, 845--862.

\bibitem[{Van~der Laan et~al.(2007)Van~der Laan, Polley and
  Hubbard}]{van2007super}
Van~der Laan, M.~J., Polley, E.~C. and Hubbard, A.~E. (2007) Super learner.
\newblock \textit{Statistical applications in genetics and molecular biology},
  \textbf{6}.

\bibitem[{Lakatos et~al.(2012)Lakatos, Golovics, David, Pandur, Erdelyi,
  Horvath, Mester, Balogh, Szipocs, Molnar et~al.}]{lakatos2012has}
Lakatos, P.~L., Golovics, P.~A., David, G., Pandur, T., Erdelyi, Z., Horvath,
  A., Mester, G., Balogh, M., Szipocs, I., Molnar, C. et~al. (2012) Has there
  been a change in the natural history of crohn's disease? surgical rates and
  medical management in a population-based inception cohort from western
  hungary between 1977--2009.
\newblock \textit{The American journal of gastroenterology}, \textbf{107}, 579.

\bibitem[{Lin et~al.(2011)Lin, Zhou, Peng and Zhou}]{lin2011selection}
Lin, H., Zhou, L., Peng, H. and Zhou, X.-H. (2011) Selection and combination of
  biomarkers using roc method for disease classification and prediction.
\newblock \textit{Canadian Journal of Statistics}, \textbf{39}, 324--343.

\bibitem[{Ludvigsson et~al.(2016)Ludvigsson, Almqvist, Bonamy, Ljung,
  Micha{\"e}lsson, Neovius, Stephansson and Ye}]{ludvigsson2016registers}
Ludvigsson, J.~F., Almqvist, C., Bonamy, A.-K.~E., Ljung, R., Micha{\"e}lsson,
  K., Neovius, M., Stephansson, O. and Ye, W. (2016) Registers of the swedish
  total population and their use in medical research.
\newblock \textit{European journal of epidemiology}, \textbf{31}, 125--136.

\bibitem[{Mogensen and Gerds(2013)}]{mogensen2013random}
Mogensen, U.~B. and Gerds, T.~A. (2013) A random forest approach for competing
  risks based on pseudo-values.
\newblock \textit{Statistics in medicine}, \textbf{32}, 3102--3114.

\bibitem[{Nicolaie et~al.(2013)Nicolaie, van Houwelingen, de~Witte and
  Putter}]{nicolaie2013dynamic}
Nicolaie, M., van Houwelingen, J., de~Witte, T. and Putter, H. (2013) Dynamic
  pseudo-observations: A robust approach to dynamic prediction in competing
  risks.
\newblock \textit{Biometrics}, \textbf{69}, 1043--1052.

\bibitem[{Ol{\'e}n et~al.(2017)Ol{\'e}n, Askling, Sachs, Frumento, Neovius,
  Smedby, Ekbom, Malmborg and Ludvigsson}]{olen2017childhood}
Ol{\'e}n, O., Askling, J., Sachs, M., Frumento, P., Neovius, M., Smedby, K.,
  Ekbom, A., Malmborg, P. and Ludvigsson, J.~F. (2017) Childhood onset
  inflammatory bowel disease and risk of cancer: a swedish nationwide cohort
  study 1964-2014.
\newblock \textit{Bmj}, \textbf{358}, j3951.

\bibitem[{Overgaard et~al.(2017)Overgaard, Parner, Pedersen
  et~al.}]{overgaard2017asymptotic}
Overgaard, M., Parner, E.~T., Pedersen, J. et~al. (2017) Asymptotic theory of
  generalized estimating equations based on jack-knife pseudo-observations.
\newblock \textit{The Annals of Statistics}, \textbf{45}, 1988--2015.

\bibitem[{Pepe et~al.(2007)Pepe, Feng, Huang, Longton, Prentice, Thompson and
  Zheng}]{pepe2007integrating}
Pepe, M.~S., Feng, Z., Huang, Y., Longton, G., Prentice, R., Thompson, I.~M.
  and Zheng, Y. (2007) Integrating the predictiveness of a marker with its
  performance as a classifier.
\newblock \textit{American journal of epidemiology}, \textbf{167}, 362--368.

\bibitem[{Polley et~al.(2017)Polley, LeDell, Kennedy and {van der
  Laan}}]{superlearner}
Polley, E., LeDell, E., Kennedy, C. and {van der Laan}, M. (2017)
  \textit{SuperLearner: Super Learner Prediction}.
\newblock \urlprefix\url{https://CRAN.R-project.org/package=SuperLearner}.
\newblock R package version 2.0-22.

\bibitem[{Ramadas et~al.(2010)Ramadas, Gunesh, Thomas, Williams and
  Hawthorne}]{ramadas2010natural}
Ramadas, A., Gunesh, S., Thomas, G., Williams, G. and Hawthorne, A. (2010)
  Natural history of crohn's disease in a population-based cohort from cardiff
  (1986--2003): a study of changes in medical treatment and surgical resection
  rates.
\newblock \textit{Gut}, \textbf{59}, 1200--1206.

\bibitem[{Saha and Heagerty(2010)}]{saha2010time}
Saha, P. and Heagerty, P. (2010) Time-dependent predictive accuracy in the
  presence of competing risks.
\newblock \textit{Biometrics}, \textbf{66}, 999--1011.

\bibitem[{Scheike et~al.(2008)Scheike, Zhang and Gerds}]{scheike2008predicting}
Scheike, T.~H., Zhang, M.-J. and Gerds, T.~A. (2008) Predicting cumulative
  incidence probability by direct binomial regression.
\newblock \textit{Biometrika}, \textbf{95}, 205--220.

\bibitem[{Zheng et~al.(2012)Zheng, Cai, Jin and Feng}]{zheng2012evaluating}
Zheng, Y., Cai, T., Jin, Y. and Feng, Z. (2012) Evaluating prognostic accuracy
  of biomarkers under competing risk.
\newblock \textit{Biometrics}, \textbf{68}, 388--396.

\end{thebibliography}

\end{document}


\section{Detailed specification of simulation study}

\subsection{Data generation mechanisms}

Let $N(\mu, \sigma)$ denote the normal distribution with mean $\mu$ and standard deviation $\sigma$. For each replicate of the simulation study, and for each of $i = 1, \ldots, 500$ independent subjects, a vector $\mathbf{X_i} = (X_{i1}, \ldots, X_{i20})$ of predictors was generated as follows: 

\begin{enumerate}
\item $X_{ij} \sim N(0, 0.1)$ for $j = 1, \ldots, 5$
\item $X_{ij} = \sum_{k = 1}^5 0.25 * X_{ik} + \varepsilon_{i}(0.1)$, for $j = 6, \ldots, 10$, where $\varepsilon_{i}(0.1) \sim N(0, 0.1)$
\item $X_{ij} = \sum_{k = 6}^{10} 0.15 * X_{ik} + \varepsilon_{i}(0.5)$, for $j = 11, \ldots, 15$, where $\varepsilon_{i}(0.5) \sim N(0, 0.5)$
\item $X_{ij} = \sum_{k = 11}^{16} 0.05 * X_{ik} + \varepsilon_{i}(0.65)$, for $j = 16, \ldots, 20$, where $\varepsilon_{i}(0.65) \sim N(0, 0.65)$
\end{enumerate}

This induces a correlation structure among the predictors. Then for each scenario the models for the scale parameter of a Weibull distribution are as follows: 

\begin{itemize}
    \item[Scenario 0:] $\gamma_{i1} = \exp{(\varepsilon_{i}(0.35))}$ where $\varepsilon_{i}(0.35) \sim N(0, 0.35)$.
    \item[Scenario A:] $\gamma_{i1} = \exp{(-2 + 2.5 * X_{i1})}$.
    \item[Scenario B:] $\gamma_{i1} = \exp{(6 + \mathbf{Z_{ib}}^T \beta_b)}$ where $\mathbf{Z_{ib}} = (X_{i1}, X_{i6}, X_{i1} * X_{i6}, b_3[X_{i1}], b_3[X_{i6}])$, and where $b_3$ denotes a b-spline basis expansion of degree 3 with 3 knots, and $\beta_b = (.75, .5, 6.1, 1.02, -2.03, 1, 2, 1, .6, .1, 3)$.
    \item[Scenario C:] $\gamma_{i1} = \exp{(\mathbf{Z_{ic}}^T \beta_c)}$, where $\mathbf{Z_{ic}} = (X_{i1}, X_{i6}, X_{i11}, X_{i16}, X_{i20}, X_{i1} * X_{i6}, \cos[X_{i11} / .1], X_{i16} * I[X_{i16} < 0])$, where $I(\cdot)$ denotes the indicator function that equals 1 if the condition is true, and 0 otherwise, and where $\beta_c = (1.1, 1.4, -2.1, -1.2, -2.3, -1.5, 6.7, .5) / 4$. 
    \item[Scenario D:] $\gamma_{i1}$ generated as in Scenario C, and additionally $\delta_{i1} = \exp((X_{i1}, X_{i6}, X_{i11}, X_{i16},X_{i20})^T \alpha_d)$ where $\alpha_d = (1.1, 1.4, -2.1, -1.2, -2.3, -1.5, 6.7, .5) / 3$.
    \item[Scenario E:] The real data example was used and bootstrapping with replacement was done to sample datasets for the simulation. 
\end{itemize}

Then, with $t_{\star} = 26.5$, $\kappa_1 = \exp(2.5 * X_{i2})$, $\kappa_2 = 2.5$, and $\gamma_2 = 3.5$, we create the rescaling terms: 

$$C_{i\gamma} = \frac{t_{\star}}{(-\log(0.8))^{-\gamma_2}}$$
and
$$C_{i\kappa} = \frac{t_{\star}}{(-\log(0.93))^{-\kappa_2}}$$ 

which are then applied as follows: 

$$
\gamma^*_{i1} = \gamma_{i1} * n^{-1}\sum \frac{C_{i\gamma}}{\gamma_{i1}}
$$
and 
$$
\kappa^*_{i1} = \kappa_{i1} * n^{-1}\sum \frac{C_{i\kappa}}{\kappa_{i1}}. 
$$

Then $Y_{i1}$ (time to surgery) is sampled from a Weibull distribution with scale $\gamma_{i1}^*$ and shape $\gamma_2$, and $Y_{i2}$ (time to death) is sampled from a Weibull distribution with scale $\kappa_{i1}^*$ and shape $\kappa_2$. The rescaling ensures that the average cumulative incidence at $t_{\star}$ is 0.2 for surgery, and 0.07 for death. 

For scenario D, censoring times $Y_{i0}$ were generated by a Weibull distribution with scale $\delta_{i1}$ and shape 1, and otherwise censoring times were uniformly distributed between the 5\% percentile of the distribution of event times and its maximum. 

For each subject, the true cumulative incidence at $t_{\star}$ for cause 1 was calculated as

\begin{equation*}
\begin{split}
P(Y_{i1} < t_{\star} \mbox{ and } Y_{i1} < Y_{i2}) = E\{P(Y_{i1} < t_{\star} \mbox{ and } Y_{i1} < x | Y_{i2} = x)\} \\
= \int_{t_\star}^\infty P(Y_{i1} < t_{\star} \mbox{ and } Y_{i1} < x) dF(x; \kappa_{i1}, \kappa_{i2}) \\
 = \int_{0}^{t_{\star}} F(x; \gamma_{i1}, \gamma_{i2}) dF(x; \kappa_{i1}, \kappa_{i2})  + 
\int_{t_{\star}}^\infty F(t_{\star}; \gamma_{i1}, \gamma_{i2}) dF(x; \kappa_{i1}, \kappa_{i2}) 
\end{split}
\end{equation*}
where $F(x; a_1, a_2)$ is the cumulative distribution function of the Weibull distribution with scale $a_1$ and shape $a_2$.

\subsection{Analysis}

For each simulated dataset, we calculated pseudo-observations at a fixed  grid of time points $\{17.5, 20, 26.5, 35\}$, where the time point of interest is 26.5, using the \texttt{pseudo} package. A validation dataset was generated independently with pseudo observations calculated only at the time point of interest. 

The SuperLearner model was run for the pseudo-observations using the entire grid of time points, with time as an additional covariate and again with only the time point of interest. The following learning algorithms from the \texttt{SuperLearner} R package were used in the library with their default settings: 

\begin{verbatim}
> SL.library
[[1]]
[1] "SL.glm"      "screen.corP"
[[2]]
[1] "SL.step"
[[3]]
[1] "SL.gam"      "screen.corP"
[[4]]
[1] "SL.ksvm"
[[5]]
[1] "SL.ranger"
[[6]]
[1] "SL.rpart"
[[7]]
[1] "SL.glmnet"
[[8]]
[1] "SL.polymars" "screen.corP"
[[9]]
[1] "xgb_200_2_0.01"
[[10]]
[1] "xgb_200_2_0.1"
[[11]]
[1] "xgb_200_2_0.2"
\end{verbatim}

The combination of learners was optimized to maximize the time varying AUC, as described in the manuscript, with a fixed $\lambda = 100$. 

The SuperLearner for the binary outcome was run using inverse probability of censoring weights estimated by the Kaplan-Meier estimate of not being censored at the minimum of the observed time and 26.5 (the time of interest). The learners in that library were the following: 

\begin{verbatim}
[[1]]
[1] "SL.glm"
[[2]]
[1] "SL.gam"
[[3]]
[1] "SL.ksvm"
[[4]]
[1] "SL.ranger"
[[5]]
[1] "SL.rpart"
[[6]]
[1] "SL.glmnet"
[[7]]
[1] "SL.polymars"
[[8]]
[1] "xgb_200_2_0.01"
[[9]]
[1] "xgb_200_2_0.1"
[[10]]
[1] "xgb_200_2_0.2"
\end{verbatim}

The weights were passed to the SuperLearner function, which passes them to the methods individually, when they are supported by that method. The combination of learners was optimized to minimize the weighted non-negative log likelihood, via \texttt{method.NNloglik}. 
The CoxBoost and survival random forests predictive models were estimated using the \texttt{CoxBoost::CoxBoost} and \texttt{randomForestsSRC::rfsrc} functions with their default settings, respectively. Predictions of the cumulative incidence function at the time point of interest were obtained by the predict methods from those packages. 

The simulations were run in R version 3.3.1 on a Windows 2008 server. 

R code for data analysis and simulations is available as an R source package at the following link. The code corresponding to what is reported in the manuscript is tagged with release date 2018-12-24. 

\href{http://github.com/sachsmc/pseupersims}{http://github.com/sachsmc/pseupersims}

\section{Description of Crohn's Disease Study Population}

\begin{table}
\caption{\label{tab1} Description of study population and potential predictors. Demographics and family characteristics. } 
\centering
\begin{tabular}{p{3cm}rrrrr}
\hline
& Censored & Surgery & Died & Emigrated &  Overall \\
& (n=7548) & (n=1397) & (n=567) & (n=93) & (n=9605) \\
\hline
\textbf{Age at onset} & & & & & \\
Mean (SD) & 40.9 (16.8) & 43.2 (18.0) & 70.9 (14.1) & 33.4 (12.0) & 42.9
(18.2) \\
Median {[}Min, Max{]} & 38 {[}18, 93{]} & 41 {[}18, 91{]} &
74 {[}19, 98{]} & 32 {[}18, 79{]} & 40 {[}18, 98{]} \\
\textbf{Male} & & & & & \\
Yes & 3553 (47.1\%) & 715 (51.2\%) & 258 (45.5\%) & 43 (46.2\%) & 4569
(47.6\%) \\
No & 3995 (52.9\%) & 682 (48.8\%) & 309 (54.5\%) & 50 (53.8\%) & 5036
(52.4\%) \\
\textbf{Inpatient} & & & & & \\
Yes & 1856 (24.6\%) & 540 (38.7\%) & 305 (53.8\%) & 24 (25.8\%) & 2725
(28.4\%) \\
No & 5692 (75.4\%) & 857 (61.3\%) & 262 (46.2\%) & 69 (74.2\%) & 6880
(71.6\%) \\
\multicolumn{3}{l}{\textbf{Highest parental education}} & & & \\
Primary school & 1211 (16.0\%) & 260 (18.6\%) & 82 (14.5\%) & 0 (0\%) &
1553 (16.2\%) \\
Secondary school & 356 (4.7\%) & 76 (5.4\%) & 7 (1.2\%) & 4 (4.3\%) &
443 (4.6\%) \\
Upper secondary school 2 years & 1888 (25.0\%) & 313 (22.4\%) & 38
(6.7\%) & 6 (6.5\%) & 2245 (23.4\%) \\
Upper secondary school 3 years & 2387 (31.6\%) & 473 (33.9\%) & 428
(75.5\%) & 64 (68.8\%) & 3352 (34.9\%) \\
University \textless{} 3 years & 733 (9.7\%) & 119 (8.5\%) & 7 (1.2\%) &
10 (10.8\%) & 869 (9.0\%) \\
University 3 years or more & 901 (11.9\%) & 148 (10.6\%) & 5 (0.9\%) & 8
(8.6\%) & 1062 (11.1\%) \\
Post graduate & 72 (1.0\%) & 8 (0.6\%) & 0 (0\%) & 1 (1.1\%) & 81
(0.8\%) \\
\multicolumn{3}{l}{\textbf{Mother born outside Sweden}} & & & \\
Yes & 919 (12.2\%) & 150 (10.7\%) & 30 (5.3\%) & 13 (14.0\%) & 1112
(11.6\%) \\
No & 6629 (87.8\%) & 1247 (89.3\%) & 537 (94.7\%) & 80 (86.0\%) & 8493
(88.4\%) \\
\multicolumn{3}{l}{\textbf{Born outside Sweden}} & & & \\
Yes & 1184 (15.7\%) & 220 (15.7\%) & 83 (14.6\%) & 62 (66.7\%) & 1549
(16.1\%) \\
No & 6364 (84.3\%) & 1177 (84.3\%) & 484 (85.4\%) & 31 (33.3\%) & 8056
(83.9\%) \\
\hline
\end{tabular}
\end{table}

\begin{table}
\caption{Description of study population and potential predictors. Medications and IBD characteristics. \label{tab1b}} 
\centering
\begin{tabular}{lrrrrr}
\hline
& Censored & Surgery & Died & Emigrated &  Overall \\
& (n=7548) & (n=1397) & (n=567) & (n=93) & (n=9605) \\
\hline
\textbf{Immunomodulators} &  1202 (15.9\%) & 143 (10.2\%) & 39 (6.9\%) & 9 (9.7\%) & 1393 (14.5\%) \\
\textbf{anti-TNF}  & 115 (1.5\%) & 13 (0.9\%) & 0 (0\%) & 0 (0\%) & 128 (1.3\%) \\
\textbf{Infliximab}  & 44 (0.6\%) & 4 (0.3\%) & 0 (0\%) & 0 (0\%) & 48 (0.5\%) \\
\textbf{Cortico steroids} & 3301 (43.7\%) & 524 (37.5\%) & 192 (33.9\%) & 20 (21.5\%) & 4037 (42.0\%) \\
\textbf{Local steroids} & 1824 (24.2\%) & 309 (22.1\%) & 77 (13.6\%) & 13 (14.0\%) & 2223 (23.1\%) \\
\textbf{Rectal ASA}  & 2181 (28.9\%) & 239 (17.1\%) & 96 (16.9\%) & 22 (23.7\%) & 2538 (26.4\%) \\
\textbf{Systemic ASA}  & 2579 (34.2\%) & 286 (20.5\%) & 113 (19.9\%) & 23 (24.7\%) & 3001 (31.2\%) \\
\textbf{Antibiotics}  & 2035 (27.0\%) & 408 (29.2\%) & 150 (26.5\%) & 11 (11.8\%) & 2604 (27.1\%) \\
\textbf{Extraintestinal Manifestations} & 172 (2.3\%) & 21 (1.5\%) & 8 (1.4\%) & 1 (1.1\%) & 202 (2.1\%) \\
\textbf{Primary Sclerosing Cholangitis}  & 29 (0.4\%) & 1 (0.1\%) & 6 (1.1\%) & 1 (1.1\%) & 37 (0.4\%) \\
\textbf{Relative has UC}  & 503 (6.7\%) & 89 (6.4\%) & 18 (3.2\%) & 3 (3.2\%) & 613 (6.4\%) \\
\textbf{Relative has CD}  & 591 (7.8\%) & 117 (8.4\%) & 17 (3.0\%) & 4 (4.3\%) & 729 (7.6\%) \\
\textbf{Stricturing/penetrating} & 6979 (92.5\%) & 1263 (90.4\%) & 515 (90.8\%) & 85 (91.4\%) & 8842 (92.1\%) \\
\textbf{Ileocecal location}  & 3236 (42.9\%) & 616 (44.1\%) & 289 (51.0\%) & 46 (49.5\%) & 4187 (43.6\%) \\
\textbf{Perianal disease}  & 457 (6.1\%) & 104 (7.4\%) & 10 (1.8\%) & 3 (3.2\%) & 574 (6.0\%) \\
\hline
\end{tabular}
\end{table}

\begin{table}
\caption{Description of study population and potential predictors. Comorbidities. \label{tab1c}} 
\centering
\begin{tabular}{lrrrrr}
\hline
& Censored & Surgery & Died & Emigrated &  Overall \\
& (n=7548) & (n=1397) & (n=567) & (n=93) & (n=9605) \\
\hline
\textbf{Diabetes} & 297 (3.9\%) & 46 (3.3\%) & 95 (16.8\%) & 1 (1.1\%) & 439 (4.6\%) \\
\textbf{Hypertension} & 642 (8.5\%) & 120 (8.6\%) & 211 (37.2\%) & 2 (2.2\%) & 975 (10.2\%) \\
\textbf{Ischemic heart disease} & 264 (3.5\%) & 50 (3.6\%) & 154 (27.2\%) & 1 (1.1\%) & 469 (4.9\%) \\
\textbf{Cerebrovascular disease} & 131 (1.7\%) & 35 (2.5\%) & 81 (14.3\%) & 0 (0\%) & 247 (2.6\%) \\
\textbf{Congestive heart disease} & 95 (1.3\%) & 23 (1.6\%) & 138 (24.3\%) & 0 (0\%) & 256 (2.7\%) \\
\textbf{COPD}  & 118 (1.6\%) & 27 (1.9\%) & 85 (15.0\%) & 0 (0\%) & 230 (2.4\%) \\
\textbf{Kidney failure} & 107 (1.4\%) & 12 (0.9\%) & 54 (9.5\%) & 0 (0\%) & 173 (1.8\%) \\
\hline
\end{tabular}
\end{table}


